\documentstyle[prb,preprint,aps]{revtex}
\draft
\tighten
\begin{document}
\title{SOLITONIC-EXCHANGE MECHANISM OF SURFACE~DIFFUSION}
\author{Oleg~M.~BRAUN
\thanks{Permanent address: Institute of Physics,
Ukrainian Academy of Sciences,
46 Science Avenue, UA-252022 Kiev, Ukraine},
Thierry~DAUXOIS\thanks{tdauxois@physique.ens-lyon.fr} and Michel~PEYRARD}
\address{Laboratoire de Physique de l'Ecole Normale Sup\'{e}rieure
de Lyon, 46 All\'{e}e d'Italie, 69364 Lyon C\'{e}dex 07, France}
\date{\today}
\maketitle
\begin{abstract}
We study surface diffusion in the framework of a generalized
Frenkel-Kontorova model with a nonconvex transverse degree
of freedom. The model describes a lattice of atoms with a
given concentration interacting by Morse-type forces, the
lattice being subjected to a two-dimensional substrate
potential which is periodic in one direction and nonconvex
(Morse) in the transverse direction. The results are used to
describe the complicated exchange-mediated diffusion
mechanism recently observed in MD simulations [J.E. Black
and Zeng-Ju Tian, Phys. Rev. Lett. {\bf 71}, 2445-2448
(1993)].
\end{abstract}
\vskip 2truecm
\pacs{PACS numbers: 03.20.+i, 46.10.+z, 68.35.Fx}
\narrowtext

\section{Introduction}
\label{intro}

Diffusion of atoms adsorbed on crystal surfaces is important
in many processes such as surface reactions and
crystal growth of artificially layered
materials~\cite{zangwill,surveys}. Besides, surface diffusion is of
considerable intrinsic interest, because experimental
results show a very rich and complicated behavior of
diffusion coefficients, especially as a function of the
atomic concentration. 

When the first adsorbed layer is complete, new
incoming atoms start to fill the second adlayer. Usually the
diffusion in the second layer follows the same laws as
that in the first layer because the adatoms of the first
monolayer play the  role  that substrate atoms played
for the diffusion of the adatoms of the first layer, i.e.\ the
first-layer adatoms create an external potential for the
second-layer atoms. However, for some adsystems the
situation may be more complicated owing to exchange of atoms
between the first and second adlayers. Such an exchange was
observed experimentally by Medvedev
{\em et al}~\cite{medvedev} for the {\bf Li}-{\bf W}(112) and {\bf
Li}-{\bf Mo}(112) adsystems, where the growth of the second
layer results in the reconstruction of the underlying first
adlayer. Moreover, recently Black and Tian~\cite{black93}  have
observed a ``complicated exchange-mediated diffusion mechanism'' in
a molecular dynamics experiment, where
an isolated {\bf Cu} adatom, which is diffusing on the {\bf
Cu}(100) surface,  may enter the first substrate
layer and create there a strain along a
close-packed row. This localized excitation moves along the row for
a distance of several lattice constants, and then the strain is
relieved by an atom in the strained row popping out and
returning to the surface. The simulation showed that this
diffusion mechanism becomes  important at  high
enough temperatures ($T\sim 900$ K for the {\bf Cu}-{\bf
Cu}(100) adsystem).

The effect of reconstructive crystal growth was
studied theoretically in ref~\onlinecite{growth} within the
framework of a generalized Frenkel-Kontorova (gFK) model
with a nonconvex transverse degree of freedom. The model describes
a chain of atoms interacting with a generalized Morse potential.
The atoms are assumed to be mobile in two directions, one is along
the surface (this is the direction along which the atoms can
diffuse), and the other one is orthogonal to the surface. They are
subjected to a two-dimensional substrate potential which is
periodic along the chain (i.e., along the surface) and has the
shape of a Morse potential in the transverse direction (orthogonal
to the surface). The concentration of atoms is characterized by the
dimensionless parameter $\theta =N/M$, the so-called coverage in
surface physics, where $N$ is the number of atoms and $M$ is the
number of minima of the external potential. This model may be
considered as a ``minimal'' model which takes into account all main
features of real adsorbed systems such as layers adsorbed on
furrowed crystal surfaces. On the other hand, it is simple enough
to be tractable analytically, at least in some aspects.
The aim of the present paper is to show that the same
model~\cite{growth} can provide a useful framework to study the
role of the atomic exchange between the first and second adlayers in
surface diffusion at coverages
$\theta \agt 1$.

In the study of surface reconstruction\onlinecite{growth}, it has
already been shown that the formation of a metastable defect
(kink) in which an atom of the second layer penetrates into the
first layer is possible in some parameter range of the gFK model.
As kinks of the first layer generally have a diffusion coefficient
$D_k$ which is much larger than the diffusion coefficient
$D_{act}$ of thermally activated atoms of the second layer,
it was speculated that the formation of such defects could have a
large influence on the overall diffusion of atoms on
crystal surfaces. The present work confirms this conjecture and
provides quantitative evaluations of the role of this
exchange-solitonic mechanism on surface diffusion.
We study a system of adsorbed atoms which
has a complete first adlayer and an extra adatom that can be in two
stable configurations. In the first one, the extra atom is in the
second layer. This state corresponds to the minimum of the total
potential energy (the gs-configuration). The second configuration
with all adatoms in the first layer corresponds to
a metastable state (the ms-configuration). In this ms-configuration
the extra atom creates a
localized solitonic excitation, the so-called kink, which usually
has a very high mobility. Thus,
even if the lifetime of the ms-configuration is
small at a nonzero temperature, the ms-configuration may
play the main role in surface diffusion, generating an unusual
temperature dependence of the mass diffusion coefficient.

The paper is organized as follows. The model is described in
Section~\ref{model}. Calculations of quasi-adiabatic
trajectories are described in Section~\ref{static}, and the
results of molecular dynamics simulation are presented
in Section~\ref{dynamic}. Section~\ref{discussion} is
devoted to theoretical estimations of different diffusion
mechanisms. The last Section~\ref{conclusion} concludes the
paper.

\section{Model}
\label{model}

We  use the generalized Frenkel-Kontorova model
introduced in ref~\onlinecite{growth}. The displacement of
an atom is characterized by two variables: $x$ describes its
motion parallel to the surface and $y$ describes its
deviation orthogonal to the substrate. The potential
perpendicular to the surface has a Morse shape
\begin{equation}
V_y(y) = \varepsilon_d \left( e^{-\gamma y}-1\right) ^2,
\label{1}
\end{equation}
which tends to the finite limit $\varepsilon_d$ (known as the
adsorption energy) when $y\rightarrow \infty$. The parameter
$\gamma$ determines the anharmonicity and it is related to
the frequency $\omega_y$ of a single-atom vibration in the
normal direction by the relation $\omega_y^2 = 2\gamma^2
\varepsilon_d /m$, $m$ being the atomic mass. It should be noticed
that the function~(\ref{1}) is nonconvex, i.e.\ it has an
inflexion point at $y=y_{inf}\equiv \gamma^{-1} \ln 2$, which
has important consequences on the properties of the
system\cite{growth}.  To model the substrate potential along the
surface, we use the deformable periodic
potential~\cite{peyrard-remoissenet},
\begin{equation}
V_x(x) = \frac{1}{2} \varepsilon_s \frac
{(1+s)^2[1-\cos (2\pi x/a_s)]}
{1+s^2-2s\cos (2\pi x/a_s)}
\label{3}
\end{equation}
as discussed in a previous work~\cite{sxy}. Here
$\varepsilon_s$ is the activation energy for diffusion of a
single atom, $a_s$ is the period of the substrate potential
along the chain, and the parameter $s$ ($|s|<1$) determines
the shape of the potential. The frequency $\omega_x$ of a
single-atom vibration along the chain is connected to the
shape parameter $s$ by $ \omega_x^2 = \omega_0^2 \left[
(1+s)/(1-s) \right] ^2 $ with $ \omega_0^2\equiv
2\pi^2\varepsilon_s / m a_s^2 $. In the following we use a
system of units where $a_s=2\pi$, $\varepsilon_s=2$ and
$m=1$, so that $\omega_0=1$.

The total potential energy of a single atom interacting with the
substrate is written as
\begin{equation}
V_{sub}(x,y) = V_x(x)e^{-\gamma'y}+V_y(y).
\label{5}
\end{equation}
The exponential factor in the first term of the right-hand
side of Eq.~(\ref{5}) takes into account the decrease of the
influence of the surface corrugation as the atoms move away
from the surface, so that $V_{sub}(x,y)\rightarrow
\varepsilon_d$ when $y\rightarrow \infty$.

To model the interaction between adatoms, we use the
generalized Morse potential
\begin{equation}
V_{int}(r) = \varepsilon_a \left\{
\frac{\beta'}{\beta-\beta'} e^{-\beta (r-r_a)} -
\frac{\beta}{\beta-\beta'} e^{-\beta' (r-r_a)}
\right\},
\label{8}
\end{equation}
where $\varepsilon_a$ is the interatomic bonding energy of a
molecule adsorbed on the surface, $r_a$ is the molecule's
equilibrium distance, and the exponents $\beta$ and $\beta'$
are related to the frequency $\omega_a$ of interatomic
vibration by the relation $ \omega_a^2 = \varepsilon_a \beta
\beta' $.

We use the same model parameters as in the
study of surface reconstruction~\onlinecite{growth}. Namely,
considering the case of metal atoms adsorbed on a metal substrate
such as, e.g., the {\bf Li}-{\bf W}(112) or the {\bf Li}-{\bf
Mo}(112) adsystems, we take $a_s = 2.73$ \AA\ which is the
distance between the wells along a furrow on the {\bf W}(112)
surface, and
$\varepsilon_s\sim 0.1$ eV, $\varepsilon_d\sim 3$ eV,
$\omega_x\alt \omega_y\sim 10^3$ cm$^{-1}$ which are typical
values for these systems. Returning to our system of units,
we get $\varepsilon_d=60$, $\gamma=0.183$, $y_{inf}=3.80$,
$\omega_x=1.5$, $\omega_y=2$, $\gamma' =2\gamma
=0.366$, and $s=0.2$. The interaction energy between two
adsorbed metal atoms usually lies within an interval
$\varepsilon_a\sim 0.1 - 0.5$ eV~\cite{BM}, or
$\varepsilon_a\sim 2 - 10$ in our system of units; we have
chosen the value $\varepsilon_a = 6$. For the exponents
$\beta$ and $\beta'$ we take $\beta=1.9$ and $\beta'=0.19$
(see a more detailed discussion of such a parameter choice
in~\onlinecite{growth}). In  the paper~\cite{growth} we
had chosen for the interatomic equilibrium distance
the value $r_a\approx 3.04$ \AA\ (the interatomic distance
in lithium metal), or $r_a=7$ in our system of units,
because we had in mind the application of the model to the
lithium film. However, we will use in the present work lower values
of $r_a$, $r_a=6.3$ and $6.4$ in order to investigate
the case when an extra atom can be inserted into the first
adlayer and exist there in a metastable state.
Although we have selected some of the parameters by comparison
with a real system, we do not claim to describe quantitatively a
concrete system of adatoms with a model which is still
oversimplified. We are interested in the {\it phenomenon} of
exchange-mediated diffusion, and this is why the parameter $r_a$
has been adjusted to allow such a phenomenon.

In the present work we study the case of a fixed
concentration of atoms. Therefore we impose periodic
boundary conditions with a fixed number $M$ of minima of
the substrate potential as well as a fixed number $N$ of
adatoms (we have used $M=16$ and $N=17$). The ground state
configuration as well as the nearest metastable states
are searched for with a standard molecular dynamics (MD)
algorithm. Namely, we are starting from an appropriate
initial configuration and allow the atoms to relax to a
nearest minimum of the total potential energy of the system.
Thus, the computer algorithm reduces to the solution of the
equations of motion which follow from the potentials~(\ref{5})
 and (\ref{8}) with an artificially introduced
viscous friction~\cite{braunKZ,growth}. In computer
simulations we can  only  include the interaction with a finite
number of neighbors. This is achieved
by introducing a cutoff distance $r^{\ast}$ (we have chosen
$r^{\ast}=5.5 \, a_s$)  and accounting only for the interactions
between the atoms separated by distances lower than $r^{\ast}$ as
usual in MD simulation.

\section{Quasi-adiabatic trajectories}
\label{static}

Adiabatic characteristics of the system  were studied  in order to
determine the potential energy surface of the model. First, we have
determined the ground state with the MD algorithm with
friction, in the same way as in the study of surface
reconstruction~\onlinecite{growth}. Then, in order to investigate mass
diffusion, we have investigated the dependence of the potential
energy upon the position of the center of mass of the system. It is
however difficult to impose a constraint on the center of mass.
Therefore, instead of determining this adiabatic trajectory, we
explore the multidimensional potential energy surface along
``quasi-adiabatic'' trajectories defined by imposing
appropriate constraints to a single atom. The
most important characteristics of the motion, which are the
positions and energies of the stationary points (the ground state,
the metastable state, and the saddle points) are calculated
correctly by this method, and we may expect to obtain the correct
shape of the adiabatic trajectory at least qualitatively with this
approach.

Since we consider the possibility that the extra atom may exist in
two stable states that differ by its distance to the substrate
(atom in the second or in the first layer), a first analysis has
been performed by constraining only the $y$ coordinate $y_j$ of
the extra atom.
We displace it up and down in the $y$ direction
(perpendicular to the surface) by small steps. At each step we
allow for the $x$ coordinate of this atom and for both coordinates
of all other atoms to adjust themselves to the new value of $y_j$.
For each relaxed state, the position $Y$ of the center of mass of
the system is calculated. Fig.~\ref{fig1} shows the
dependence of the system energy upon $Y$.
The ground state
configuration corresponds to a $Y$-position of the kink
around 2.8, while the metastable state corresponds to a much
lower value of $Y$. These pictures allow us to calculate
$\varepsilon_{ms}$, the difference in energies of the
metastable and groundstate. The results are listed in
Table~\ref{table1}. One can notice that $\varepsilon_{ms}$
depends very much on $r_a$; the metastable
state is much more likely to be relevant in the case (a) $r_a =
6.3$ which has a lower
$\varepsilon_{ms}$ than in case (b) $r_a = 6.4$. The two stable
states are separated by a  barrier $\varepsilon_{barrier}$.

Around each value of $Y$ corresponding to a stable state, we have
then determined the potential energy as a function of the
displacement along $x$ of the coordinate $x_j$ of the extra atom.
In this case $x_j$ is constrained to a given value but $y_j$ is
allowed to relax to its equilibrium value as well as the $x$ and $y$
coordinates of all the other atoms. The $X$ coordinate of the
center of mass is again calculated for each relaxed state. The
results are presented in Fig.~\ref{fig2} for the case of a
quasi-adiabatic trajectory starting from the ground state, and in
Fig.~\ref{fig3} for a trajectory starting from the metastable
state. In this case, the structure of the metastable defect is such
that the extra atom does not play a specific role and cannot be
identified unambiguously. This has no consequence on the mass
diffusion, but in the algorithm one has to chose the atom that will
be constrained to a given $x_j$. The choice has been done in the
same way as in ref~\cite{braunKZ}. Figure~\ref{fig2} shows
that, for the translation of the center of mass along $x$ in the
case of an extra atom in the second layer,  there is no metastable
state and only a barrier to overcome. The diffusion in the
$X$-direction will therefore be an activated process and we call
$\varepsilon_{act}$ the difference between the unstable maximum
state and the stable one. As attested by the results presented in
Table~\ref{table1},  $\varepsilon_{act}$ decreases when
$r_a$ increases. More precise consequences for the physics
will be explained in the following section. Figure \ref{fig3} and
Table~\ref{table1} show that the barrier $\varepsilon_{pn}$ for
the $X$ translation of the metastable state is extremely low
compared to $\varepsilon_{act}$.

In order to complete the picture of the potential energy
surface, in a second series of calculations we artificially
moved the same atom~$j$, but now both coordinates
$x_j$ and $y_j$ were constrained, while the other atoms were allowed
to relax to the minimum of the system energy. The atom $j$ was moved
to scan the $x$ and $y$ directions as in a TV sweep. The results
allow to draw the full picture of the energy $E$ of the system as
a function of $X$ and $Y$. They are presented in
Fig.~\ref{fig4}a as a contour map and in
Fig.~\ref{fig4}b as a three-dimensional surface (we show
the results  only for $r_a=6.3$ , because the
case $r_a=6.4$ looks qualitatively similar).

>From the energy surface of Fig.~\ref{fig4}
and the quantitative results of Table~\ref{table1}, one can
predict the general behavior of the system. Starting from
the ground state in the second layer, the extra adatom may
overcome the barrier
$\varepsilon_{act}$ and directly jump to the nearest
neighboring site in the second layer, or it may overcome
the barrier $\varepsilon_{barrier}$ and form the
metastable state in the first adlayer that will move
practically without barriers. Therefore during the lifetime
of the metastable state the local compression of the first layer
may move for a long distance, and when the system
returns back to the ground state an atom  arises in the
second layer in a site which may be far away from the
initial position of the extra atom. Such an exchange-mediated
diffusion should clearly be the favorable diffusion pathway in the
case
$r_a=6.3$ because, in this case, the activation
barrier for the transition to the metastable state is lower than
the activation energy in the second layer,
$\varepsilon_{barrier} <
\varepsilon_{act}$. On the other hand, in the case $r_a=6.4$
we have the opposite inequality, $\varepsilon_{barrier} >
\varepsilon_{act}$, and direct jumps to a nearest site of the
second layer should be the most probable diffusion path. However,
even in this case, transitions to the metastable state may take
place and, owing to the high kink mobility, these transitions may
lead to a remarkable contribution to the total
diffusion coefficient. This aspect is discussed in the following
section.

\section{Molecular dynamics simulations.}
\label{dynamic}

The predictions presented above are simply based on static
configuration energies and collective dynamical effects could enter
in a non trivial way to affect the diffusion. Thus these predictions
must be checked with full molecular dynamics simulations
of the system. Temperature effects are introduced through a
standard Langevin approach. The hamiltonian equation of motion of
each atom of mass $m$ is completed by a friction term with a
damping coefficient $m \eta$ and a random force with a $\delta$
correlation function of factor $2 m \eta k_B T$.
Starting from the ground state configuration a first time interval
$t_{th}$  is allowed to reach the
thermal equilibrium state. Then, during a time interval $t_{run}$
we save the dependencies $X(t)$, $Y(t)$, which define
the position of the localized excitation
and $y_{max}(t)$  which defines the maximum $y$-coordinate
of the atoms at time $t$. This last quantity allows us to
determine
if the system is in the ground state or in the metastable state.

A first step of the analysis has been devoted to the properties of
the metastable at non zero temperature. In principle the position
$Y$ of the center of mass of the system should tell us whether the
system of adatoms is in the ground state or the metastable state.
However, in practice, measures of $Y$ do not answer this
question because the histogram for the probability $P(Y)$ that $Y$
takes a given value shows only one maximum corresponding to the
ground state configuration. This does not mean that the
metastable is never excited, but, because of the importance
of the fluctuations it is not possible to distinguish the two stable
states, characterized by the shift between the two layers of only
one atom (among N=17 atoms):  the increase of $Y$ is too small.
The information on the state of the system can be deduced from
$y_{max}$ because this quantity takes a large value if an
atom is in the second layer (i.e. when the system is in the ground
state). The histogram of the probability
$P(y_{max})$ does exhibit two
well-defined maxima corresponding to the ground state and
metastable state configurations up to temperatures above
$T=2$ (see Fig.~\ref{fig5}a). From this result, it is possible to
define an effective free energy of the system as
\begin{equation}
F_{eff}=-k_B T \ln P(y_{max}).
\label{Veff}
\end{equation}
which is plotted in Fig.~\ref{fig6} at different temperatures. The
difference in free energy between  the metastable state and the
ground state (called $\varepsilon_{ms}$ in Sec.~~\ref{static}), as
well as the height  $\varepsilon_{barrier}$ of the barrier between
the ground state and the metastable state are approximately equal
to the values deduced from quasi-adiabatic calculations at
$T=0$. The relative vertical positions of the curves are not
significant, because they depend on the reference in energy
that we chose for each temperature. Note, however, that the
free energy of the metastable state with respect to the
ground state free energy, $\varepsilon_{ms}(T)$, increases with
temperature from the value
$0.6$ at $T=0.3$ to $1.2$ at $T=0.9$. This effect may be
understood by noticing that the variation of this difference
in free energy is related to the entropy of both
states as follows:
\begin{equation}
{\partial \varepsilon_{ms}(T)\over \partial
T}={\partial\left[F_{eff}(y_{max}^{ms})-
F_{eff}(y_{max}^{gs})\right] \over
\partial T}= S_{gs}-S_{ms}.
\end{equation}
Since this variation is positive according to the numerical
result, it means that  the entropy of the ground state is higher
than that of the metastable state. This results is reasonable
because the atom isolated in the second layer has more space
available to move than when it is in the first layer, and moreover
its interaction energy with its neighbors and with the substrate
(because of the
$\exp(-\gamma'y)$~factor) is weaker.

Integrating the peaks of the histogram of $P(y_{max})$, we can
obtain the occupation numbers of the two states. The density of
state for the ground state,
$\rho_{gs}$, is calculated by considering the states on the right
of the minimum value between the two peaks. The density of state
for the metastable state is then $\rho_{ms}=1-\rho_{gs}$. In
Fig.~\ref{fig5}b they are shown as functions of the
temperature, and compared with the functions
\begin{equation}
\rho_{gs}^{(0)}=\left( 1+e^{-\varepsilon_{ms}/k_B T} \right)^{-1}
\;\;\;\mbox{and}\;\;\;
\rho_{ms}^{(0)}=1-\rho_{gs}^{(0)},
\label{rho}
\end{equation}
which would be deduced from the quasi-adiabatic results without
taking  into account the variation of the energy of
the metastable state with temperature. A significant deviation
between the numerical values and the estimation provided by the
quasi-adiabatic results shows up only in the high temperature
range.

The second set of studies has been devoted to a direct
determination of the mass diffusion coefficient from the
thermalized molecular dynamics simulation. The result is deduced
from a series of $k$ simulations at each temperature
($k=20$), each one extending over the time interval $t_{run}$. From
simulation $i$ we extract the diffusion coefficient using
\begin{equation}
D_i=\lim_{t \to \infty}{\left<[X(t_0+t)-X(t_0)]^2\right>\over 2t}
\end{equation}
(where $t_0+t$ is lower than $t_{run}$).
 Then we
average over the
$k$ simulations. The results are presented in Figs.~\ref{fig7}
and~\ref{fig8}, and discussed in the following section.

Besides the total diffusion coefficient, we calculate also
the kink diffusion coefficient (i.e. the diffusion
coefficient of the system in the metastable state). For
this, we started from the metastable state and monitor
the maximum $y_{max}$ coordinate of the atoms: when an atom
escapes from the first adlayer, we stop the run. A reliable
numerical estimation of the kink diffusion coefficient is only
possible at low temperature when the metastable has a sufficient
lifetime. The results  are presented in Fig.~\ref{fig9}.

\section{Discussion}
\label{discussion}

Let us now come to the main point,
an analytical estimate of the role of kinks in the process
of surface diffusion. As shown above,
the system may evolve from one GS configuration
to an another GS configuration
with the atom occupying a neighboring site in the second adlayer
by two pathways, either a direct atomic jump over the
barrier
$\varepsilon_{act}$ or through a two-stage mechanism
involving the metastable state in the first adlayer.
For such a two-way process the total diffusion coefficient
can be calculated by the method described by
Kutner and Sosnowska~\cite{Kutner75}.
The result is trivial:
\begin{equation}
D=\rho_{gs} D_{act}+\rho_{ms} D_{kink},
\label{D}
\end{equation}
where the occupation numbers $\rho_{gs}$ and $\rho_{ms}$
were defined above.

The hopping diffusion coefficient $D_{act}$ may be estimated
with the help of the seminal result  of the Kramers
theory~\cite{Hanggi}
\begin{equation}
D_{act} \approx
\left(\sqrt{\omega_b^2+{\eta^2\over
4}}-{\eta\over2}\right){\omega^{\ast}\over
\omega_b} \frac{a_s^2}{2 \pi}
\exp \left( -\frac{\varepsilon_{act}}{k_B T} \right).
\label{Da} 
\end{equation}
where $\omega^{\ast}$
(respectively $\omega_b$) is
the frequency of the linearized oscillations of the system near
the ground (resp. unstable)
state. As shown on Fig.~\ref{fig2}, the potential
energy for the motion over the barrier is well approximated by
a sinusoidal function
$V(x)=\varepsilon_{act}{\left[1+\cos(2\pi x/ a_s)\right]/2}$,
so that $\omega^{\ast}=\omega_b= \sqrt{\varepsilon_{act} /2 m}$.
Finally, as $a_s=2\pi$ and $m=1$ with our units, we get
\begin{equation}
D_{act} \approx
\left(\sqrt{{\varepsilon_{act} \over 2 }+{\eta^2\over
4}}-{\eta\over2}\right){2 \pi}
\exp \left( -\frac{\varepsilon_{act}}{k_B T} \right),
\label{Dabis}
\end{equation}

To estimate the diffusion coefficient $D_{kink}$ of the metastable
state, recall that the atom inserted into the chain
creates a local distortion that can be considered as a kink
in the Frenkel-Kontorova model. In the continuum approximation the
equation of motion reduces to the exactly integrable
sine-Gordon (SG) equation. This limit, in which a defect would
propagate freely as a soliton in the system, can be used as a
starting point for a perturbative analysis of the influence of the
discreteness of the lattice. When the continuous translational
invariance is broken, one finds that the kink experiences a
periodic potential, with the periodicity of the lattice, which is
known as the Peierls-Nabarro barrier in dislocation theory. A
precise evaluation of the PN barrier turns out to involve subtle
effects, particularly when discreteness effects are
weak\cite{flachwillis}, but a recent analysis has derived accurate
results in this case\cite{FLACH}. The basic results is that the
shape of the kink can be obtained by solving a Klein-Gordon
equation obtained by replacing the actual substrate potential
$V(x)$ by an effective potential
\begin{equation}
V_{eff}(x) = V(x) - {1 \over 24 g} \bigg[ V^{\prime} (x) \bigg]^2
\label{veff}
\end{equation}
where $g = a_s^2 V''_{int}(a_s) /( 2\pi^2\varepsilon_s)
= V''_{int}(a_s)$ is the elastic constant of the atomic chain.
This expression allows us to derive the value of the mass of
the kink which is given by the general
expression for a Klein Gordon model,
\begin{equation}
m_{kink} = {m \over a_s^2 \sqrt{g}} \int_{0}^{a_s}
\sqrt{2 V_{eff}(x)} dx \; .
\label{massgl}
\end{equation}
This formula gives the usual expression
$m_{kink}= 8\, m / (a_s^2 \sqrt{g})$ in the sine-Gordon case if,
instead of the effective potential, we use the actual potential
$V(x) = [1 - \cos(2 \pi x / a_s)]$. For the values of $g$ which
are obtained with our parameters (see Table~\ref{table2}), the
correction to the SG value due to discreteness is small, giving for
instance
$m_{kink}=7.95\, m / (a_s^2 \sqrt{g})$ for the case $r_a = 6.3$.
The proper treatment of discreteness is much more important for
the evaluation of the amplitude of the Peierls potential which is
obtained as\cite{FLACH}
\begin{equation}
\varepsilon_{pn} \approx 712.26\ \varepsilon_{s}\
 g\ \exp (-\pi^2 \sqrt g)\quad.
\label{pn}
\end{equation}
Table~\ref{table2} gives the values of $g$, $m_{kink}$ and
$\varepsilon_{pn}$ which result from
the interaction potential~(\ref{8}) with $r_a = 6.3$ and $r_a=6.4$.
It shows that the amplitude of the PN potential is negligible
with respect to the barrier $\varepsilon_{act}$ corresponding to
the translation of the extra atom in the second layer. This is
in agreement of the numerical results of Sect.~\ref{static}.
Moreover, this value $\varepsilon_{pn} \ll k_B T$
at the temperature that we consider, shows that the kink motion is
not thermally activated. Its diffusion coefficient can be derived
from the formula for a free diffusion
\begin{equation}
D_{kink}={k_B T\over m_{kink}\ \eta_{kink}},
\label{Dk}
\end{equation}
where $\eta_{kink}$ is an effective viscous friction for kink
motion. At $T=0$ $\eta_{kink}$ is simply given by $\eta_{kink}
\approx
\eta$, where $\eta$ is the viscous friction for the motion of an
isolated adatom due to energy exchange with the substrate.
The value of $\eta$ is usually~\cite{fric} about $\eta \sim
\omega_0/10$.
But at nonzero temperature $\eta_{kink}$ depends on $T$
because of the coupling of the lattice phonons with the
highly nonlinear core of the kink.
The simplest form for this dependence is~\cite{marchesoni}
\begin{equation}
\eta_{kink} \approx \eta+\alpha T,
\label{etak}
\end{equation}
where $\alpha$ is a coefficient which cannot be obtained
analytically but could be obtained experimentally
as in the case of copper~\cite{ALERSTHOMPSON}.

In our analysis, we treat $m_{kink}$ and $\alpha$ as
adjustable parameters. They are chosen by
fitting  with Eq.~(\ref{Dk}) the temperature evolution of $D_{kink}$
determined  by the MD simulations with
$\eta=0.1$, $r_a=6.3$. The parameters $m_{kink}=0.0886$
and $\alpha=0.155$ obtained in this particular
case  are also valid for the case $\eta=0.3$ as shown
from Fig.~\ref{fig9}.
Moreover, the expression (\ref{Dk}) with the {\it same} parameters
describes the simulation results for the diffusion in the first adlayer
for the case $r_a=6.4$ as well. This rather good agreement between
the numerical results and the analytical estimation of $D_{kink}$
shows that, although the SG description may seem rather crude for
the generalized FK model, it provides a good basis for analysis.
This is confirmed by the comparison between the
fitted value of
$m_{kink}$ and the theoretical value given by
Eq.~(\ref{massgl}): the fitted value is smaller than the
theoretical one, but the order of magnitude of $m_{kink}$ is
however correctly given by the discrete SG calculation.

Having obtained analytical estimates for the two diffusion
coefficients $D_{act}$ and $D_{kink}$, we are now in a position to
estimate the role of the kink diffusion in the general process of
atomic diffusion in the adsorbed layer. The role of both
mechanism is well illustrated on Fig.~\ref{fig7}. First one can
observe that the temperature dependence of the total diffusion
coefficient $D$ deduced from the MD simulations is well reproduced
by Eq.~(\ref{D}). It should be noticed that, at this level the fit
is obtained {\it without} adjustable parameters since our
parameters have been deduced from the study of $D_{kink}$ alone
and the density of states $\rho_{gs}$ and $\rho_{ms}$ are the
measured quantities. The good agreement points out that the
diffusion mechanism that we propose, involving two different
processes, provides a correct description of diffusion as it can
be observed in MD simulations of the generalized FK model. It is
also important to notice that {\it both mechanisms} are essential
to reproduce the numerical results. If one would consider only
one of the two processes, i.e. assume that
$\rho_{gs}=1$ or $\rho_{ms}=1$ , the diffusion
coefficient would evolve versus $T$ along the dash-dotted or
dotted lines of Fig.~\ref{fig7}. The theoretical value of
$D$ would disagree completely with the MD results.

Fig.~\ref{fig8} summarizes the results by providing a comparison
between the temperature dependence of $D$ observed by MD and the
theoretical value of Eq.~(\ref{D}) for $r_a = 6.3$ and $r_a = 6.4$
and two values of the damping coefficient $\eta$. This figure
shows that all the numerical results are well described by the
two-process diffusion with only two adjustable parameters
$\alpha$ and $m_{kink}$. The results show also that, even in the
case $r_a=6.4$, for which the direct jumps to a nearest adsite in
the second layer has
to overcome a much lower barrier than the barrier for the transition
to the metastable state, the solitonic-exchange mechanism
involving the metastable state still brings a crucial
contribution to the total diffusion coefficient $D$. This is due
to the extremely high mobility of the metastable state.

\section{Conclusion}
\label{conclusion}

Following our results, one can distinguish three different
diffusion mechanisms for atoms adsorbed on a crystalline surface.

The first one is the {\sl conventional diffusion} (sometimes
called hopping diffusion) which involves only one atomic layer.
In can be an individual process
when an adatom directly jumps from one adsorption site to a nearest
adsite. The diffusion coefficient in this case follows the
Arrhenius law $D(T)=A
\exp \left( -\varepsilon_{act} /k_B T \right)$ with the
preexponential factor being independent (or weakly
dependent) on $T$. At high coverages $\theta\alt 1$ (as well
as for $\theta\alt 2$, etc) owing to the interaction between
the adatoms, the activation energy $\varepsilon_{act}$
decreases with respect to the height of the original
substrate potential $\varepsilon_s$. For a strong
interatomic interaction the motion takes a collective
(concerted) character and may be described with the help of
the kink concept~\cite{braunKZ,devil}. In this case the
activation energy $\varepsilon_{act}$ for the chemical
diffusion becomes the Peierls-Nabarro barrier
$\varepsilon_{pn}$ which is significantly lower than the individual
activation energy. Therefore the process is generally no longer
thermally activated and $D(T)$ is approximately given by the free
diffusion formula $D(T) \approx
k_B T / m_{kink} \eta_{kink}$.

While for the conventional diffusion the underlying adatoms
in the first adlayer play a passive role creating the
effective external potential for the diffusion of atoms in the
second adlayer, for the {\sl exchange diffusion} mechanism
this role becomes active. Two types of  mechanisms can be considered.
The   ``conventional'' (or ``one-step'') exchange diffusion
mechanism has been known for a long time. It occurs when an
adatom
$A$ from the second layer ``pushes out'' an adatom $B$ from its
regular position in the first adlayer and occupies the
free site so created. The new configuration has only a
single adatom ($B$) in the second adlayer. The intermediate
configuration with  both adatoms $A$ and $B$ inserted into
the first adlayer is
unstable. It corresponds to a saddle point in the potential energy
surface. The elementary diffusion step takes a short time $\sim
10^{-13}$ sec. The diffusion coefficient for the one-step exchange
mechanism should follow the same Arrhenius law as for the
conventional hopping diffusion with the activation energy
corresponding to the saddle state. Exchange diffusion may be
important for coverages $\theta \agt 1$, when the first adlayer is
complete and the second one starts to grow.

The new diffusion mechanism studied in the present work,
is a {\sl two-step exchange} diffusion mechanism.
The main difference with one-step exchange
diffusion is that the configuration with the adatom inserted into
the first adlayer, now corresponds to a {\em metastable} state
instead of the unstable (saddle) state.
Because this metastable state corresponds to a kink configuration
which is characterized by a very high mobility, the
mechanism may be called as the ``exchange-solitonic''
mechanism of surface diffusion. For this mechanism the
diffusion follows an Arrhenius law too, but now the
activation energy corresponds to the difference in energies
between the metastable state and the ground state (and not
to the barrier for the transition from the second adlayer to
the first one), while the preexponent factor is determined
by the kink diffusion coefficient and essentially depends on
temperature. The mass diffusion coefficient in this
situation is described by the expression~(\ref{D}) where
$D_{act}$ and $D_{kink}$ are given by Eqs.~(\ref{Dabis}) and
(\ref{Dk}).

We considered above the situation where an atom is diffusing
in the second adlayer over a filled first adlayer. It is
clear that the same situation can occur for adatoms of the
first layer when the adatoms are the same as the atoms of
the substrate. Indeed, in this case
the top-layer substrate atoms play the same role as
that of the first-layer adatoms in the former case. In particular
the exchange diffusion mechanism was observed for the first time
when De Lorenzi and Jacucci~\cite{Lorenzi85} investigated
by the MD method the self-diffusion of {\bf Na} atoms
adsorbed on the surface of a {\bf Na} metal crystal.
They found that the exchange mechanism is responsible for
the diffusion across the rows on the furrowed crystal surface.
This conclusion was later confirmed by the field ion microscope
technique for the diffusion of atoms adsorbed on the transition
metal surfaces~\cite{zangwill,surveys}.

It seems possible and would be
interesting to find and investigate the exchange-solitonic
mechanism of surface diffusion experimentally with the
field-ion microscope or STM technique. It may be predicted
that such a diffusion is to be expected for coverages $\theta
\sim a_s /r_a$.

\acknowledgments

The authors thank Thierry Cretegny for helpful discussions.
Part of this work has been supported by the NATO grant LG920236.



\begin{table}
\caption{Parameters of quasi-adiabatic trajectories for two
values of the equilibrium distance $r_a$ of the interatomic
potential. $\varepsilon_{ms}$ is the difference in energies
of the metastable and ground state configurations,
$\varepsilon_{barrier}$ is the activation barrier for the
transition from the gs-configuration to the metastable
state, $\varepsilon_{act}$ is the activation energy for the
hopping diffusion, and $\varepsilon_{pn}$ is the barrier for
kink motion parallel to the surface.}
\vskip 1truecm

\begin{tabular}{ccccccccc}
parameter                       & $r_a=6.3$       & $r_a=6.4$  \\
\tableline
$\varepsilon_{ms}$              & 0.755           & 3.014 \\
$\varepsilon_{barrier}$         & 1.752           & 3.476 \\
$\varepsilon_{act}$             & 2.364           & 2.055 \\
$\varepsilon_{pn}$              & $4.18\cdot 10^{-6}$
                                                  & $6.39\cdot 10^{-6}$ \\
\end{tabular}
\label{table1}
\end{table}

\begin{table}
\caption{Parameters corresponding to the kink (metastable state):
$g$ is the elastic constant,
$m_{kink}$ is the kink mass, and
$\varepsilon_{pn(theory)}$ is the amplitude of the Peierls-Nabarro barrier.}

\vskip 1truecm
\begin{tabular}{ccccccccc}
parameter                       & $r_a=6.3$       & $r_a=6.4$  \\
\tableline
$g$                             & 2.243           & 2.759 \\
$m_{kink}$                      & 0.1344           & 0.1214 \\
$\varepsilon_{pn(theory)}$      & $1.22\cdot 10^{-3}$
                                                  & $2.98\cdot 10^{-4}$
\end{tabular}
\label{table2}
\end{table}


\begin{figure}
\caption{Change of the energy $\Delta E$ with respect to the
ground state as function of $Y$ for a displacement along the
quasi-adiabatic trajectory  perpendicular to the surface for (a)
$r_a=6.3$ and (b) $r_a=6.4$.}
\label{fig1}
\end{figure}

\begin{figure}
\caption{Change of the energy $\Delta E$ with respect to the
ground state  as function of $X$ for a quasi-adiabatic
displacement parallel to the surface starting from the ground state
for (a) $r_a=6.3$ and (b) $r_a=6.4$.}
\label{fig2}
\end{figure}

\begin{figure}
\caption{Change of the energy $\Delta E$
as function of $X$ for a quasi-adiabatic
displacement parallel to the surface starting from the metastable state
for (a) $r_a=6.3$ and (b) $r_a=6.4$.}
\label{fig3}
\end{figure}

\begin{figure}
\caption{Potential energy surface for the case $r_a=6.3$.
The $X$-$Y$ contour map is plotted in figure (a) while
figure (b) presents a three-dimensional plot.}
\label{fig4}
\end{figure}

\begin{figure}
\caption{Simulation results for $r_a=6.3$ and $\eta=0.1$.
(a) Probability of the maximum of the $y$ position
for different temperatures.
The solid line corresponds to $T=0.3$,
the dotted line corresponds to $T=0.5$,
the dashed line corresponds to $T=0.7$, and
the dot-dashed line corresponds to $T=0.9$.
(b) Density of states versus temperature.
The solid line and stars describe the simulation results,
whereas the dashed line corresponds to the estimation
(\protect \ref{rho}).}
\label{fig5}
\end{figure}

\begin{figure}
\caption{The effective free energy (\protect \ref{Veff}) as function of
the maximum of the $y$ atomic positions as follows from the results
of Fig.~\protect \ref{fig5}.
The solid line corresponds to $T=0.3$,
the dotted line corresponds to $T=0.5$,
the dashed line corresponds to $T=0.7$, and
the dot-dashed line corresponds to $T=0.9$.}
\label{fig6}
\end{figure}

\begin{figure}
\caption{Diffusion coefficient versus temperature
for $r_a=6.3$ and $\eta=0.1$.
The diamonds correspond to the simulation results.
The error bars are computed numerically with 20 simulations.
The dotted curve corresponds to $D_{kink}$
and the dash-dotted curve corresponds to $D_{act}$.
The stars and the solid curve describe the estimation result
$D=\rho_{gs} D_{act} + \rho_{ms} D_{kink}$,
where $\rho_{gs}$ and $\rho_{ms}$ were taken from the simulation
results of Fig.~\protect \ref{fig5}b.}
\label{fig7}
\end{figure}

\begin{figure}
\caption{Comparison between the MD results and the theoretical
values of the diffusion coefficient in four cases. The symbols
correspond to the simulation results, whereas the solid curves, to
the theoretical ones. The diamonds describe the case $r_a=6.3$ and
$\eta=0.1$, the triangles, the case $r_a=6.3$ and $\eta=0.3$,
the squares correspond to the case $r_a=6.4$ and $\eta=0.1$, and
the stars, to the case $r_a=6.4$ and $\eta=0.3$.}
\label{fig8}
\end{figure}

\begin{figure}
\caption{The kink diffusion coefficient versus
temperature for $r_a=6.3$.
The symbols correspond to the simulation results
and the lines describe the estimated results of
Eq.~(\protect \ref{Dk}).
The stars and the dotted curve correspond to $\eta=0.1$,
the diamonds and the dashed curve, to $\eta=0.3$.
The parameters $m_{kink}$ and $\alpha$ were obtained by fitting
the expression~(\protect \ref{Dk}) to the simulation data
for the case $\eta=0.1$,
and then the same values were used for $\eta=0.3$.}
\label{fig9}
\end{figure}
\end{document}